\documentclass{article}

\usepackage[a4paper]{anysize}
\usepackage{amsthm,amsmath,graphicx}

\title{The boundary value contact problem of electroelasticity for
  piecewise-homogeneous piezo-electric plate with elastic inclusion
  and cut}

\author{Nugzar Shavlakadze\footnote{Iv. Javakhishvili Tbilisi State
    University A.Razmadze Mathematical Institute, Tamarashvili str. 6,
    0177, Tbilisi, Georgia e-mail:nusha@rmi.ge}\and
  Nana Odishelidze\footnote{Department of Computer Sciences
    Faculty of Exact and Natural Sciences,Iv.Javakhishvili Tbilisi
    State University, 2, University st., 0143,Georgia.  Tfn:
    99532304784. E-mail: nana\_georgiana@yahoo.com} \and
  Francisco Criado-Aldeanueva\footnote{Department of Applied Physics
    II, Polytechnic School, Malaga University, Campus Teatinos, s/n
    (29071), Spain. Tfn: +34952132849. E-mail:
    fcaldeanueva@ctima.uma.es}}

\renewcommand{\Re}{\text{Re}\,}
\renewcommand{\Im}{\text{Im}\,}

\begin{document}
\maketitle

\begin{abstract}
  A contact problem of the theory of electroelasticity for
  piecewise-homogeneous plate of piezo-electric material with infinite
  cut and elastic finite inclusion of variable bending rigidity is
  considered.  By using methods of the theory of analytic function,
  the problem is reduced to a system of singular integro-differential
  equation with fixed singularity. Using an integral transformation a
  Riemann problem is obtained, the solution of which is presented in
  explicit form.


\end{abstract}

\section{Introduction}

Exact or approximate solutions of static contact problems for
different domains, reinforced with elastic mountings, thin inclusions
or patches of variable stiffness were obtained earlier, and the
behavior of the contact stresses at the ends of the contact line have
been investigated as a function of the law of variation of the
geometrical and physical parameters of these components
\cite{1,6,7,5,3,11,2,12,13,14}. In homogeneity problems are addressed
in \cite{27,25,26,30,29,28,9,10,8}. The first fundamental problem for
a piecewise-homogeneous plane was solved when a crack of finite length
arrives at the interface of two bodies at the right angle \cite{15},
and also a similar problem for a piecewise-homogeneous plane when
acted upon symmetrical normal stresses at the crack sides
\cite{16,17}, as well as the contact problems for
piecewise-homogeneous planes with a semi-infinite and finite inclusion
\cite{18,19}.

\section{Problem statement and its solution}

We will consider a piecewise-homogeneous plate of piezo-electric
material, weakened with infinite crack and reinforced with a finite
inclusion(beam) as an electrode by a normal force of intensity
$p_0(x)$. Let us assume that $p_0(x)$ is bounded function on the
segment.  The normal stresses $q_0(x)$ and the electric potential are
given at the edges of the crack.

\paragraph{The problem}
consists of determining the expansion of the cut and the jump $p(x)$
of normal contact stresses along the contact line and of establishing
their behavior in the neighborhood of the ends of the inclusion.  It
is formulated as follows: suppose an elastic body $S$ occupies the
plane of complex variable $z=x+iy$, which contains, along the section
$l_1=(0,1)$ an elastic inclusion and an infinite cut along the
half-axis $l_2=(-\infty,0)$ and consists of two half-planes of
dissimilar piezo-electric materials
\begin{equation*}
  S^{(1)} = \{z | \Re z > 0, \quad z \not\in [0,1]\}, \qquad
  S^{(2)} = \{z | \Re z > 0, \quad z \not\in (-\infty,0)\}, \qquad
\end{equation*}
joined along the $OY$ axis. Quantities and functions, referred to the
half-plane $S^{(k)}$, will be denoted by the subscript $k$ ($k=1,2$),
while the boundary values of the other functions on the upper and
lower sides of the patch will by denoted by a plus or minus sign,
respectively.

\begin{figure}[htbp]
  \centering
  \includegraphics{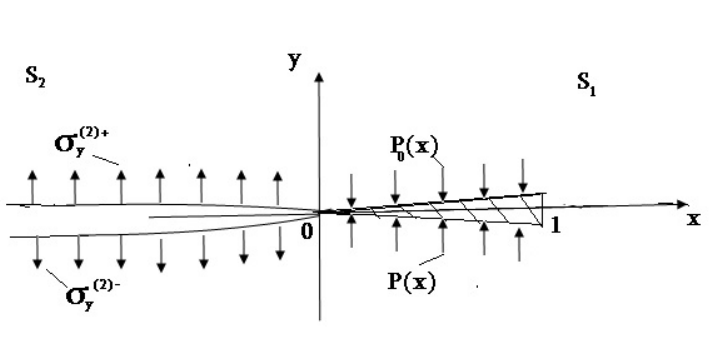}
  \caption{Statement of the problem}\label{fig:1}
\end{figure}

In conditions of plane deformation on plate acts at infinity the
homogeneous fields of mechanical and electrical stresses:
$\sigma_{11}^\infty$, $\sigma_{33}^\infty$, $\tau_{13}^\infty$,
$E_1^\infty = E_3^\infty = 0$. At the boundary of the inclusion
electrical field's potential is $\varphi_1^+ = \varphi_1^- = 0$ and at
the boundary of the crack $\sigma_y^{(2)+}(x) =
\sigma_y^{(2)-}(x) = q_0(x)$, $\tau_{xy}^{(2)+}(x) =
\tau_{xy}^{(2)-}(x) = 0$, $\varphi_2^+ = \varphi_2^- =
\varphi_2(x)$. (see Fig. \ref{fig:1})
 
According to the equilibrium equation of inclusions elements and
Hooke's law we have
\begin{equation}\label{eq:1}
  \frac{d^2}{dx^2} D(x) \frac{d^2 \nu^{(1)}(x)}{dx^2} = p_0(x) - p(x), \qquad
  0 < x < 1
\end{equation}
and the equilibrium equation of the inclusion has the form
\begin{equation}\label{eq:2}
  \int_0^1 [p(t) - p_0(t)] \, dt = 0, \qquad
  \int_0^1 t [p(t) - p_0(t)] \, dt = 0
\end{equation}
where $\nu^{(1)}(x)$ is the vertical displacement of inclusion points
and $p(x)$ is the jump of tangential contact stresses to be
determined. $D(x) = E_1(x) h_1^3(x) / (1-\nu_1^2)$ with $E_1(x)$,
$h_1(x)$, $\nu_1$ the modulus of elasticity , thickness and Poisson's
coefficient of the inclusions material, respectively.

At the boundary of the crack we have
\begin{equation}\label{eq:3}
  \sigma_y^{(2)+}(x) + \sigma_y^{(2)-}(x) = 2 q_0(x),  \qquad x < 0
\end{equation}
In the interface of two materials we have
\begin{equation}\label{eq:4}
  \begin{aligned}
    \sigma_x^{(1)} &= \sigma_x^{(2)}, &
    \tau_{xy}^{(1)} &= \tau_{xy}^{(2)}, &
    u^{(1)} &= u^{(2)},\\
    \nu^{(1)} &= \nu^{(2)}, &
    E_y^{(1)} &= E_y^{(2)}, &
    D_x^{(1)} &= D_x^{(2)},
  \end{aligned}
\end{equation}
where $\sigma_x^{(j)}$, $\tau_{xy}^{(j)}$ are stress components;
$u^{(j)}$, $\nu^{(j)}$ are displacements components, $E_y^{(j)}$ and
$D_x^{(j)}$ are components of vectors of electrical stress and of
electrical inductive ($j=1,2$).

In the plane $XOY$ for stress function $\varphi_1^{(j)}$ and
electrical field's potential $\varphi_2^{(j)}$ we obtain the system of
differential equations \cite{20}:
\begin{equation}\label{eq:5}
  l_{11}^{(j)}\varphi_1^{(j)} + l_{12}^{(j)}\varphi_2^{(j)} = 0, \qquad
  l_{12}^{(j)}\varphi_1^{(j)} + l_{22}^{(j)}\varphi_2^{(j)} = 0  
\end{equation}
where
\begin{equation*}
  \begin{aligned}
    l_{11}^{(j)} &= a_{10}^{(j)} \partial_1^4 + a_{12}^{(j)} \partial_1^2\partial_2^2 + a_{14}^{(j)} \partial_2^4, &
    \partial_1 &= \frac \partial{\partial x}, &
    \partial_2 &= \frac \partial{\partial y}\\
    l_{12}^{(j)} &= l_{21}^{(j)} = a_{21}^{(j)} \partial_1^2 \partial_2 + a_{23}^{(j)} \partial_2^3, &
    l_{22}^{(j)} &= a_{20}^{(j)} \partial_1^2 + a_{22}^{(j)} \partial_2^2, &
    a_{10}^{(j)} &= s_{33}^{(j)} - (s_{13}^{(j)})^2(s_{11}^{(j)})^{-1}\\
    a_{12}^{(j)} &= s_{44}^{(j)} + 2 s_{13}^{(j)} (1 - s_{12}^{(j)} (s_{11}^{(j)})^{-1}),&
    a_{14}^{(j)} &= s_{11}^{(j)} - (s_{12}^{(j)})^2 (s_{11}^{(j)})^{-1},\\
    a_{21}^{(j)} &= s_{13}^{(j)} d_{13}^{(j)} (s_{11}^{(j)})^{-1} - d_{33}^{(j)} + d_{15}^{(j)}, &
    a_{23}^{(j)} &= d_{13}^{(j)} \left(s_{12}^{(j)} (s_{11}^{(j)})^{-1} - 1\right),\\
    a_{20}^{(j)} &= \epsilon_{11}^{(j)}, &
    a_{22}^{(j)} &= \epsilon_{33}^{(j)} -(d_{13}^{(j)})^2 (s_{11}^{(j)})^{-1}, &
    j &= 1,2.
  \end{aligned}
\end{equation*}
where $s_{nk}^{(j)}$, $d_{nk}^{(j)}$, $\epsilon_{nk}^{(j)}$ are
elastic tractability, piezoelectric modules and dielectric constants,
respectively.

General solutions of equations (\ref{eq:5}) are represented using three analytical functions
\begin{equation}\label{eq:6}
  \begin{aligned}
    \varphi_1^{(j)} &= 2 \Re \sum_{k=1}^3 \gamma_k^{(j)} \int \Phi_k^{(j)} (z_k^{(j)})\, dz_k^{(j)}, &
    \varphi_2^{(j)} &= - 2 \Re \sum_{k=1}^3 \lambda_k^{(j)} \Phi_k^{(j)} (z_k^{(j)})\\
    z_k^{(j)} &= x + \mu_k^{(j)} y, \qquad
    \mu_{3+k}^{(j)} = \overline{\mu}_k^{(j)}, &
    \gamma_k^{(j)} &= a_{20}^{(j)} + a_{22}^{(j)} (\mu_k^{(j)})^2,\\
    \lambda_k^{(j)} &= a_{21}^{(j)} \mu_k^{(j)} + a_{23}^{(j)} (\mu_k^{(j)})^3
  \end{aligned}
\end{equation}
$\mu_k^{(j)}$ are roots of characteristic equations:
\begin{equation*}
  c_0^{(j)}(\mu^{(j)})^6 + c_1^{(j)} (\mu^{(j)})^4 + c_2^{(j)} (\mu^{(j)})^2 + c_3^{(j)} = 0, \qquad
  k=1,2,3, \quad j=1,2,
\end{equation*}
where                                                              
\begin{equation*}
  \begin{aligned}
    c_0^{(j)} &= a_{14}^{(j)} a_{22}^{(j)} - (a_{23}^{(j)})^2, &
    c_1^{(j)} &= a_{12}^{(j)} a_{22}^{(j)} + a_{14}^{(j)} a_{20}^{(j)} - 2 a_{21}^{(j)} a_{23}^{(j)}, &
    c_2^{(j)} &= a_{10}^{(j)} a_{22}^{(j)} + a_{12}^{(j)} a_{20}^{(j)} - a_{21}^{(j)},\\
    c_3^{(j)} &= a_{10}^{(j)} a_{20}^{(j)}, &
    \Im \mu_k^{(j)} &\neq 0
  \end{aligned}
\end{equation*}
Using formulas (\ref{eq:6}) we obtain representation for stress
component, displacements, vectors of electrical stress and of
electrical inductive:
\begin{equation*}
  \begin{aligned}
    \sigma_x^{(j)} &= 2 \Re \sum_{k=1}^3 \gamma_k^{(j)} (\mu_k^{(j)})^2 {\Phi'}_k^{(j)} (z_k^{(j)}), &
    \sigma_y^{(j)} &= 2 \Re \sum_{k=1}^3 \gamma_k^{(j)} {\Phi'}_k^{(j)} (z_k^{(j)}),\\
    \tau_{xy}^{(j)} &= - 2 \Re \sum_{k=1}^3 \gamma_k^{(j)} \mu_k^{(j)} {\Phi'}_k^{(j)} (z_k^{(j)}), &
    u^{(j)} &= 2 \Re \sum_{k=1}^3 p_k^{(j)} \Phi_k^{(j)} (z_k^{(j)}),\\
    \nu^{(j)} &= 2 \Re \sum_{k=1}^3 q_k^{(j)} \Phi_k^{(j)} (z_k^{(j)}), & 
    E_x^{(j)} &= 2 \Re \sum_{k=1}^3 \lambda_k^{(j)} {\Phi'}_k^{(j)} (z_k^{(j)}),\\
    E_y^{(j)} &= 2 \Re \sum_{k=1}^3 \lambda_k^{(j)} \mu_k^{(j)} {\Phi'}_k^{(j)} (z_k^{(j)}), &
    D_x^{(j)} &= 2 \Re \sum_{k=1}^3 r_k^{(j)} \mu_k^{(j)} {\Phi'}_k^{(j)} (z_k^{(j)}),\\
    D_y^{(j)} &= - \Re \sum_{k=1}^3 r_k^{(j)} {\Phi'}_k^{(j)} (z_k^{(j)}),
  \end{aligned}
\end{equation*}
where
\begin{equation*}
  \begin{aligned}
    p_k^{(j)}
    &= a_{14}^{(j)} \gamma_k^{(j)} (\mu_k^{(j)})^2
    + \frac 12 (a_{12}^{(j)} - s_{44}^{(j)}) \gamma_k^{(j)}
    - a_{23}^{(j)} \lambda_k^{(j)} \mu_k^{(j)},\\
    q_k^{(j)}
    &= \frac 12 (a_{12}^{(j)} - s_{44}^{(j)}) \gamma_k^{(j)} \mu_k^{(j)}
    + a_{10}^{(j)} \gamma_k^{(j)} (\mu_k^{(j)})^{-1}
    - (a_{21}^{(j)} - d_{15}^{(j)}) \lambda_k^{(j)},\\
    r_k^{(j)}
    &= a_{20}^{(j)} \lambda_k^{(j)} (\mu_k^{(j)})^{-1}
    - d_{15}^{(j)} \gamma_k^{(j)}.
  \end{aligned}
\end{equation*}

Introducing the notation $H_k^{(j)}(x) = [{\Phi'}_k^{(j)}(x)]^+ -
[{\Phi'}_k^{(j)}(x)]^-$, ($k=1,2,3$, $j=1,2$) the boundary value
conditions
\begin{equation*}
  \begin{aligned}
    \sigma_y^{(1)+} - \sigma_y^{(1)-} &= p(x), &
    \tau_{xy}^{(1)+} - \tau_{xy}^{(1)-} &= 0\\
    \left(\frac{\partial u^{(1)}}{\partial x}\right)^+ - 
    \left(\frac{\partial u^{(1)}}{\partial x}\right)^- &= 0, &
    \left(\frac{\partial \nu^{(1)}}{\partial x}\right)^+ - 
    \left(\frac{\partial \nu^{(1)}}{\partial x}\right)^- &= 0,\\
    E_x^{(1)+}(x) &= E_x^{(1)-}(x), &
    D_y^{(1)+}(x) &= D_y^{(1)-}(x), &
    x \in l_1\\
    \\
    \sigma_y^{(2)+} - \sigma_y^{(2)-} &= 0, &
    \tau_{xy}^{(2)+} - \tau_{xy}^{(2)-} &= 0\\
    \left(\frac{\partial u^{(2)}}{\partial x}\right)^+ - 
    \left(\frac{\partial u^{(2)}}{\partial x}\right)^- &= 0, &
    \left(\frac{\partial \nu^{(2)}}{\partial x}\right)^+ - 
    \left(\frac{\partial \nu^{(2)}}{\partial x}\right)^- &= 2 \left(\frac{\partial \nu}{\partial x}\right)^+ \equiv 2 f(x),\\
    E_x^{(2)+}(x) &= E_x^{(2)-}(x) = -\frac{\partial \varphi_2}{\partial x}, &
    D_y^{(2)+}(x) &= D_y^{(2)-}(x), &
    x \in l_2
  \end{aligned}
\end{equation*}
can be represented using three analytical functions :
\begin{subequations}
  \begin{equation}\label{eq:7a}
    \Re \sum_{k=1}^3 r_k^{(1)} H_k^{(1)}(x) = 0, \qquad
    \Re \sum_{k=1}^3 \gamma_k^{(1)} H_k^{(1)}(x) = \frac{p(x)}2, \qquad
    \Re \sum_{k=1}^3 p_k^{(1)} H_k^{(1)}(x) = 0, \qquad
    x\in l_1
  \end{equation}
  \begin{equation}\label{eq:7b}
    \Re \sum_{k=1}^3 \gamma_k^{(1)} \mu_k^{(1)} H_k^{(1)}(x) = 0, \qquad
    \Re \sum_{k=1}^3 \lambda_k^{(1)} H_k^{(1)}(x) = 0, \qquad
    \Re \sum_{k=1}^3 q_k^{(1)} H_k^{(1)}(x) = 0, \qquad
    x\in l_1
  \end{equation}
\end{subequations}
\begin{subequations}
  \begin{equation}\label{eq:8a}
    \Re \sum_{k=1}^3 r_k^{(2)} H_k^{(2)}(x) = 0, \qquad
    \Re \sum_{k=1}^3 \gamma_k^{(2)} H_k^{(2)}(x) = 0, \qquad
    \Re \sum_{k=1}^3 p_k^{(2)} H_k^{(2)}(x) = 0, \qquad
    x\in l_2
  \end{equation}
  \begin{equation}\label{eq:8b}
    \Re \sum_{k=1}^3 \gamma_k^{(2)} \mu_k^{(2)} H_k^{(2)}(x) = 0, \qquad
    \Re \sum_{k=1}^3 \lambda_k^{(2)} H_k^{(2)}(x) = 0, \qquad
    \Re \sum_{k=1}^3 q_k^{(2)} H_k^{(2)}(x) = f(x), \qquad
    x\in l_2
  \end{equation}
\end{subequations}
Without loss of generality, assume that $\mu_k^{(j)} = i
\beta_k^{(j)}$, $k=1,2,3$, $j=1,2$ \cite{21,22}.

Then one obtains
\begin{align}
  [{\Phi'}_k^{(1)}(x)]^+ - [{\Phi'}_k^{(1)}(x)]^- &=  \frac{\Delta_{2k}^{(1)}}{2 \Delta_0^{(1)}} p(x), & k=1,2,3, \quad 0 < x < 1\label{eq:9}\\
  [{\Phi'}_k^{(2)}(x)]^+ - [{\Phi'}_k^{(2)}(x)]^- &=  \frac{\Delta_{3k}^{(2)}}{\Delta_0^{(2)}} i f(x), & k=1,2,3, \quad x < 0\label{eq:10}
\end{align}
where $\Delta_0^{(1)} \neq 0$ is determinant of system (\ref{eq:7a}),
$\Delta_{2k}^{(1)}$ is corresponding algebraic additions,
$-i\Delta_0^{(2)} \neq 0$ is determinant of system (\ref{eq:8b}),
$\Delta_{3k}^{(2)}$ is corresponding algebraic additions.

The solutions of the problems of linear conjugation with boundary
conditions (\ref{eq:9}) and (\ref{eq:10}) are represented in the form
\begin{equation}\label{eq:11}
  \begin{aligned}
    {\Phi'}_k^{(1)} (z_k^{(1)})
    &= \frac{\Delta_{2k}^{(1)}}{4\pi i \Delta_0^{(1)}} \int_0^1 \frac{p(t)\, dt}{t-z_k^{(1)}}
    + W_k^{(1)} (z_k^{(1)}) \equiv \Psi_k^{(1)} (z_k^{(1)}) + W_k^{(1)}(z_k^{(1)}), & z_k^{(1)} \in S_k^{(1)}\\
    {\Phi'}_k^{(2)} (z_k^{(2)})
    &= \frac{\Delta_{3k}^{(2)}}{2\pi \Delta_0^{(2)}} \int_{-\infty}^0 \frac{f(t)\, dt}{t-z_k^{(2)}}
    + W_k^{(2)} (z_k^{(2)}) \equiv \Psi_k^{(2)} (z_k^{(2)}) + W_k^{(2)}(z_k^{(2)}), & z_k^{(2)} \in S_k^{(2)}
  \end{aligned}
\end{equation}
where $W_k^{(j)}(z_k^{(j)})$ are analytic functions in the half-plates
$S_k^{(j)}$, $k=1,2,3$, $j=1,2$.

To determinate the analytic functions $W_k^{(1)}(z_k^{(1)})$ we obtain
the following equations from (\ref{eq:4}) (the boundary condition on
the interface of two materials)
\begin{equation*}
  \begin{aligned}
    \sum_{k=1}^3 \gamma_k^{(1)} (\beta_k^{(1)})^2 M_k(t_k^{(1)}) &= \sum_{k=1}^3 \gamma_k^{(2)} (\beta_k^{(2)})^2 M_k(t_k^{(2)}), &
    \sum_{k=1}^3 \gamma_k^{(1)} i \beta_k^{(1)} \widetilde M_k(t_k^{(1)}) &= \sum_{k=1}^3 \gamma_k^{(2)} i \beta_k^{(2)} \widetilde M_k(t_k^{(2)}),\\
    \sum_{k=1}^3 \lambda_k^{(1)} i \beta_k^{(1)} M_k(t_k^{(1)}) &= \sum_{k=1}^3 \lambda_k^{(2)} i \beta_k^{(2)} M_k(t_k^{(2)}), &
    \sum_{k=1}^3 p_k^{(1)} i \beta_k^{(1)} \widetilde M_k(t_k^{(1)}) &= \sum_{k=1}^3 p_k^{(2)} i \beta_k^{(2)} \widetilde M_k(t_k^{(2)}),\\
    \sum_{k=1}^3 q_k^{(1)} i \beta_k^{(1)} M_k(t_k^{(1)}) &= \sum_{k=1}^3 q_k^{(2)} i \beta_k^{(2)} M_k(t_k^{(2)}), &
    \sum_{k=1}^3 r_k^{(1)} i \beta_k^{(1)} \widetilde M_k(t_k^{(1)}) &= \sum_{k=1}^3 r_k^{(2)} i \beta_k^{(2)} \widetilde M_k(t_k^{(2)})
  \end{aligned}
\end{equation*}
where
\begin{equation*}
  \begin{aligned}
    M_k(t_k^{(j)})
    &= W_k^{(j)} (t_k^{(1)}) + \overline{W}_k^{(j)} (\overline t_k^{(j)})
    + \Psi_k^{(j)} (t_k^{(j)}) + \overline{\Psi}_k^{(j)}(\overline t_k^{(j)})\\
    \widetilde M_k(t_k^{(j)})
    &= W_k^{(j)} (t_k^{(j)}) - \overline{W}_k^{(j)} (\overline t_k^{(j)})
    + \Psi_k^{(j)} (t_k^{(j)}) - \overline{\Psi}_k^{(j)}(\overline t_k^{(j)})\\
    t_k^{(j)} &= i \beta_k^{(j)} y, \qquad k=1,2,3, \quad j=1,2.
  \end{aligned}
\end{equation*}
After multiplication of the obtained expressions by $\frac 1{2\pi i}
\frac {dt}{t-z}$, $t=iy$, $z=x+iy$ and integrating along axis $OY$, by
using Cauchy's theorem and formula, we obtain the system of algebraic
equations with respect to $W_k^{(1)}(\beta_k^{(1)}z)$, $\overline
W_k^{(2)}(-\beta_k^{(2)}z)$ ($k=1,2,3$)

\begin{multline*}
  \sum_{k=1}^3 [\gamma_k^{(1)} (\beta_k^{(1)})^2 W_k^{(1)} (\beta_k^{(1)} z)
    - \gamma_k^{(2)} (\beta_k^{(2)})^2 \overline W_k^{(2)} (-\beta_k^{(2)} z)]\\
  =
  - \sum_{k=1}^3 \gamma_k^{(1)} (\beta_k^{(1)})^2 \overline \Psi_k^{(1)} (-\beta_k^{(1)} z)
  + \sum_{k=1}^3 \gamma_k^{(2)} (\beta_k^{(2)})^2 \overline \Psi_k^{(2)} (\beta_k^{(2)} z)
\end{multline*}
\begin{multline*}
  \sum_{k=1}^3 [\gamma_k^{(1)} i \beta_k^{(1)} W_k^{(1)} (\beta_k^{(1)} z)
    + \gamma_k^{(2)} i \beta_k^{(2)} \overline W_k^{(2)} (-\beta_k^{(2)} z)]\\
  =
  - \sum_{k=1}^3 \gamma_k^{(1)} i \beta_k^{(1)} \overline \Psi_k^{(1)} (-\beta_k^{(1)} z)
  + \sum_{k=1}^3 \gamma_k^{(2)} i \beta_k^{(2)} \Psi_k^{(2)} (\beta_k^{(2)} z)
\end{multline*}
\begin{multline*}
  \sum_{k=1}^3 [\lambda_k^{(1)} i \beta_k^{(1)} W_k^{(1)} (\beta_k^{(1)} z)
    - \lambda_k^{(2)} i \beta_k^{(2)} \overline W_k^{(2)} (-\beta_k^{(2)} z)]\\
  =
  - \sum_{k=1}^3 \lambda_k^{(1)} i \beta_k^{(1)} \overline \Psi_k^{(1)} (-\beta_k^{(1)} z)
  + \sum_{k=1}^3 \lambda_k^{(2)} i \beta_k^{(2)}  \Psi_k^{(2)} (\beta_k^{(2)} z)
\end{multline*}
\begin{multline*}
  \sum_{k=1}^3 [p_k^{(1)} i \beta_k^{(1)} W_k^{(1)} (\beta_k^{(1)} z)
    - p_k^{(2)} i \beta_k^{(2)} \overline W_k^{(2)} (-\beta_k^{(2)} z)]\\
  =
  \sum_{k=1}^3 p_k^{(1)} i \beta_k^{(1)} \overline \Psi_k^{(1)} (-\beta_k^{(1)} z)
  + \sum_{k=1}^3 p_k^{(2)} i \beta_k^{(2)} \Psi_k^{(2)} (\beta_k^{(2)} z)
\end{multline*}
\begin{multline*}
  \sum_{k=1}^3 [q_k^{(1)} i \beta_k^{(1)} W_k^{(1)} (\beta_k^{(1)} z)
    - q_k^{(2)} i \beta_k^{(2)} \overline W_k^{(2)} (-\beta_k^{(2)} z)]\\
  =
  - \sum_{k=1}^3 q_k^{(1)} i \beta_k^{(1)} \overline \Psi_k^{(1)} (-\beta_k^{(1)} z)
  + \sum_{k=1}^3 q_k^{(2)} i \beta_k^{(2)} \Psi_k^{(2)} (\beta_k^{(2)} z)
\end{multline*}
\begin{multline*}
  \sum_{k=1}^3 [r_k^{(1)} i \beta_k^{(1)} W_k^{(1)} (\beta_k^{(1)} z)
    + r_k^{(2)} i \beta_k^{(2)} \overline W_k^{(2)} (-\beta_k^{(2)} z)]\\
  =
  \sum_{k=1}^3 r_k^{(1)} i \beta_k^{(1)} \overline \Psi_k^{(1)} (-\beta_k^{(1)} z)
  + \sum_{k=1}^3 r_k^{(2)} i \beta_k^{(2)} \Psi_k^{(2)} (\beta_k^{(2)} z)
\end{multline*}

Solving this system we obtain
\begin{equation}\label{eq:12}
  \begin{aligned}
    W_1^{(1)}(z_1^{(1)})
    &= \sum_{k=1}^3 A_k^{(1)} \overline\Psi_k^{(1)}\left(-\frac{\beta_k^{(1)}}{\beta_1^{(1)}}z_1^{(1)}\right)
    + \sum_{k=1}^3  A_k^{(2)} \Psi_k^{(2)}\left(-\frac{\beta_k^{(2)}}{\beta_1^{(1)}}z_1^{(1)}\right)\\
    W_2^{(1)}(z_2^{(1)})
    &= \sum_{k=1}^3 B_k^{(1)} \overline\Psi_k^{(1)}\left(-\frac{\beta_k^{(1)}}{\beta_2^{(1)}}z_2^{(1)}\right)
    + \sum_{k=1}^3  B_k^{(2)} \Psi_k^{(2)}\left(-\frac{\beta_k^{(2)}}{\beta_2^{(1)}}z_2^{(1)}\right)\\
    W_3^{(1)}(z_3^{(1)})
    &= \sum_{k=1}^3 C_k^{(1)} \overline\Psi_k^{(1)}\left(-\frac{\beta_k^{(1)}}{\beta_3^{(1)}}z_3^{(1)}\right)
    + \sum_{k=1}^3  C_k^{(2)} \Psi_k^{(2)}\left(-\frac{\beta_k^{(2)}}{\beta_3^{(1)}}z_3^{(1)}\right)\\
    \overline W_1^{(2)}(-z_1^{(2)})
    &= \sum_{k=1}^3 D_k^{(1)} \overline\Psi_k^{(1)}\left(-\frac{\beta_k^{(1)}}{\beta_1^{(2)}}z_1^{(2)}\right)
    + \sum_{k=1}^3  D_k^{(2)} \Psi_k^{(2)}\left(\frac{\beta_k^{(2)}}{\beta_1^{(2)}}z_1^{(2)}\right)\\
    \overline W_2^{(2)}(-z_2^{(2)})
    &= \sum_{k=1}^3 E_k^{(1)} \overline\Psi_k^{(1)}\left(-\frac{\beta_k^{(1)}}{\beta_2^{(2)}}z_2^{(2)}\right)
    + \sum_{k=1}^3  E_k^{(2)} \Psi_k^{(2)}\left(\frac{\beta_k^{(2)}}{\beta_2^{(2)}}z_2^{(2)}\right)\\
    \overline W_3^{(2)}(-z_3^{(2)})
    &= \sum_{k=1}^3 F_k^{(1)} \overline\Psi_k^{(1)}\left(-\frac{\beta_k^{(1)}}{\beta_3^{(2)}}z_3^{(2)}\right)
    + \sum_{k=1}^3  F_k^{(2)} \Psi_k^{(2)}\left(\frac{\beta_k^{(2)}}{\beta_3^{(2)}}z_3^{(2)}\right)
  \end{aligned}
\end{equation}
where
\begin{equation*}
  \begin{aligned}
    A_k^{(j)} =& \left(
    (-1)^j \gamma_k^{(j)} (\beta_k^{(j)})^2 \widetilde A_{11}
    + i \gamma_k^{(j)} \beta_k^{(j)} \widetilde A_{21}
    + (-1)^j i \lambda_k^{(j)} \beta_k^{(j)} \widetilde A_{31}
    + p_k^{(j)} i \beta_k^{(j)} \widetilde A_{41}\right.\\
    &\left.+ (-1)^{(j)} q_k^{(j)} i \beta_k^{(j)} \widetilde A_{51}
    + r_k^{(j)} i \beta_k^{(j)} \widetilde A_{61}
    \right) / \widetilde \Delta\\
    B_k^{(j)} =& \left(
    \gamma_k^{(j)} (\beta_k^{(j)})^2 \widetilde A_{12}
    + i \gamma_k^{(j)} \beta_k^{(j)} \widetilde A_{22}
    + i \lambda_k^{(j)} \beta_k^{(j)} \widetilde A_{32}
    + p_k^{(j)} i \beta_k^{(j)} \widetilde A_{42}
    + q_k^{(j)} i \beta_k^{(j)} \widetilde A_{52}
    + r_k^{(j)} i \beta_k^{(j)} \widetilde A_{62}
    \right) / \widetilde \Delta\\
    C_k^{(j)} =& \left(
    (-1)^j \gamma_k^{(j)} (\beta_k^{(j)})^2 \widetilde A_{13}
    + i \gamma_k^{(j)} \beta_k^{(j)} \widetilde A_{23}
    + (-1)^j i \lambda_k^{(j)} \beta_k^{(j)} \widetilde A_{33}
    + p_k^{(j)} i \beta_k^{(j)} \widetilde A_{43}
    + (-1)^j q_k^{(j)} i \beta_k^{(j)} \widetilde A_{53}\right.\\
    &\left.+ r_k^{(j)} i \beta_k^{(j)} \widetilde A_{63}
    \right) / \widetilde \Delta\\
    D_k^{(j)} =& \left(
    \gamma_k^{(j)} (\beta_k^{(j)})^2 \widetilde A_{14}
    + i \gamma_k^{(j)} \beta_k^{(j)} \widetilde A_{24}
    + i \lambda_k^{(j)} \beta_k^{(j)} \widetilde A_{34}
    + p_k^{(j)} i \beta_k^{(j)} \widetilde A_{44}
    + q_k^{(j)} i \beta_k^{(j)} \widetilde A_{54}
    + r_k^{(j)} i \beta_k^{(j)} \widetilde A_{64}
    \right) / \widetilde \Delta\\
    E_k^{(j)} =& \left(
    (-1)^j \gamma_k^{(j)} (\beta_k^{(j)})^2 \widetilde A_{15}
    + i \gamma_k^{(j)} \beta_k^{(j)} \widetilde A_{25}
    + (-1)^j i \lambda_k^{(j)} \beta_k^{(j)} \widetilde A_{35}
    + p_k^{(j)} i \beta_k^{(j)} \widetilde A_{45}
    + (-1)^j q_k^{(j)} i \beta_k^{(j)} \widetilde A_{55}\right.\\
    &\left.+ r_k^{(j)} i \beta_k^{(j)} \widetilde A_{65}
    \right) / \widetilde \Delta\\
    F_k^{(j)} =& \left(
    \gamma_k^{(j)} (\beta_k^{(j)})^2 \widetilde A_{16}
    + i \gamma_k^{(j)} \beta_k^{(j)} \widetilde A_{26}
    + i \lambda_k^{(j)} \beta_k^{(j)} \widetilde A_{36}
    + p_k^{(j)} i \beta_k^{(j)} \widetilde A_{46}
    + q_k^{(j)} i \beta_k^{(j)} \widetilde A_{56}
    + r_k^{(j)} i \beta_k^{(j)} \widetilde A_{66}
    \right) / \widetilde \Delta, \\
    k &= 1,2,3, \quad j= 1,2
  \end{aligned}
\end{equation*}

\begin{equation*}
  \widetilde\Delta =
  - \prod\limits_{k=1}^3 \beta_k^{(1)}  \beta_k^{(2)}
  \begin{vmatrix}
    - i \gamma_1^{(1)} \beta_1^{(1)} & - i \gamma_2^{(1)} \beta_2^{(1)} & - i \gamma_3^{(1)} \beta_3^{(1)} & 
    i \beta_1^{(2)} \gamma_1^{(2)} & i \beta_2^{(2)} \gamma_2^{(2)} & i \beta_3^{(2)} \gamma_3^{(2)}\\
    \gamma_1^{(1)} & \gamma_2^{(1)} & \gamma_3^{(1)} & \gamma_1^{(2)} & \gamma_2^{(2)} & \gamma_3^{(2)}\\
    \lambda_1^{(1)} & \lambda_2^{(1)} & \lambda_3^{(1)} & -\lambda_1^{(2)} & -\lambda_2^{(2)} & -\lambda_3^{(2)}\\
    p_1^{(1)} & p_2^{(1)} & p_3^{(1)} & p_1^{(2)} & p_2^{(2)} & p_3^{(2)}\\
    q_1^{(1)} & q_2^{(1)} & q_3^{(1)} & -q_1^{(2)} & -q_2^{(2)} & -q_3^{(2)}\\
    r_1^{(1)} & r_2^{(1)} & r_3^{(1)} & r_1^{(2)} & r_2^{(2)} & r_3^{(2)}
  \end{vmatrix}\neq 0
\end{equation*}
where $\widetilde A_{ij}$ are corresponding algebraic additions.

Since $\Re\widetilde \Delta = 0$, we have $\Im \{A_k,B_k,C_k\} = 0$,
$k=1,2,3$.

On the bases of conditions (\ref{eq:2}-\ref{eq:3}) and formulas
(\ref{eq:11}-\ref{eq:12}), we obtain the following system of singular
integro-differential equations
\begin{align}
  \frac{d^2}{dx^2} D(x) \frac{d}{dx} \left(
  \lambda_1 \int_0^1 \frac{p(t)dt}{t-x}
  +\lambda_2 \int_0^1 \frac{p(t)dt}{t+x}
  + \int_0^1 R_1(t,x) p(t) dt
  + \int_{-\infty}^0 R_2(t,x) f(t) dt
  \right)
  &= p_0(x) - p(x), & x&\in l_1\label{eq:13}\\
  \lambda_3 \int_{-\infty}^0 \frac{f(t)dt}{t-x}
  + \lambda_4 \int_{-\infty}^0 \frac{f(t)dt}{t+x}
  + \int_0^1 R_3(t,x) p(t) dt
  + \int_{-\infty}^0 R_4(t,x) f(t) dt
  &= q_0(x), &
  x&\in l_2\label{eq:14}
\end{align}
where
\begin{gather*}
  R_1(t,x) = \sum_{m\neq n = 1}^3 \frac{\omega_{mn}}{\beta_m^{(1)} t + \beta_n^{(1)}x}, \qquad
  R_2(t,x) = \sum_{m\neq n = 1}^3 \frac{\alpha_{mn}}{\beta_m^{(1)} t - \beta_n^{(2)}x},\\
  \lambda_1 = \sum_{k=1}^3 \frac{i q_k^{(1)} \Delta_k^{(1)}}{2 \pi \Delta_0^{(1)}}, \qquad
  \lambda_2 = i \frac{q_1^{(1)} A_1^{(1)} \Delta_1^{(1)} + q_2^{(1)} B_2^{(1)} \Delta^{(1)}_2 + q_3^{(1)} C_3^{(1)} \Delta_3^{(1)}}{2 \pi \Delta_0^{(1)}},\qquad
  \omega_{12} = \beta_1^{(1)} \frac{i q_1^{(1)} A_2^{(1)} \Delta_2^{(1)}}{2 \pi \Delta_0^{(1)}}\\
  \omega_{21} = \beta_2^{(1)} \frac{i q_2^{(1)} B_1^{(1)} \Delta_1^{(1)}}{2 \pi \Delta_0^{(1)}},\qquad
  \omega_{13} = \beta_1^{(1)} \frac{i q_1^{(1)} A_3^{(1)} \Delta_3^{(1)}}{2 \pi \Delta_0^{(1)}}, \qquad
  \omega_{31} = \beta_3^{(1)} \frac{i q_3^{(1)} C_1^{(1)} \Delta_1^{(1)}}{2 \pi \Delta_0^{(1)}}\\
  \omega_{23} = \beta_2^{(1)} \frac{i q_2^{(1)} B_3^{(1)} \Delta_3^{(1)}}{2 \pi \Delta_0^{(1)}},\qquad
  \omega_{32} = \beta_3^{(1)} \frac{i q_3^{(1)} C_2^{(1)} \Delta_2^{(1)}}{2 \pi \Delta_0^{(1)}}\\
  \alpha_{1n} = \frac{-2 i q_1^{(1)} A_1^{(n)} \Delta_n^{(2)} \beta_1^{(1)}}{\pi \Delta_0^{(2)}}, \quad
  \alpha_{2n} = \frac{-2 i q_2^{(1)} B_1^{(n)} \Delta_n^{(2)} \beta_2^{(1)}}{\pi \Delta_0^{(2)}}, \quad
  \alpha_{3n} = \frac{-2 i q_3^{(1)} C_1^{(n)} \Delta_n^{(2)} \beta_3^{(1)}}{\pi \Delta_0^{(2)}}, \qquad
  n= 1,2,3.\\
  R_3(t,x) = \sum_{m\neq n=1}^3 \frac{r_{mn}}{\beta_m^{(2)} t - \beta_n^{(1)} x}, \qquad
  R_4(t,x) = \sum_{m\neq n = 1}^3 \frac{q_{mn}}{\beta_m^{(2)} t + \beta_n^{(2)} x},\\
  \lambda_3 = -2\sum_{k=1}^3 \frac{\gamma_k^{(2)} \Delta_k^{(2)}}{\pi \Delta_0^{(2)}}, \qquad
  \lambda_4 = -2 \frac{\gamma_1^{(2)} D_1^{(2)} \Delta_1^{(2)} + \gamma_2^{(2)} E_2^{(2)} \Delta_2^{(2)} + \gamma_3^{(2)} F_3^{(2)} \Delta_3^{(2)}}{\pi\Delta_0^{(2)}}\\
  r_{1n} = \frac{\gamma_1^{(2)} D_n^{(1)} \Delta_n^{(2)} \beta_1^{(2)}}{2\pi \Delta_0^{(1)}}, \quad
  r_{2n} = \frac{\gamma_2^{(2)} E_n^{(1)} \Delta_n^{(1)} \beta_2^{(2)}}{2\pi \Delta_0^{(1)}}, \quad
  r_{3n} = \frac{\gamma_3^{(2)} F_n^{(1)} \Delta_n^{(1)} \beta_3^{(2)}}{2\pi \Delta_0^{(1)}}, \qquad
  n=1,2,3.\\
  q_{12} = \beta_1^{(2)} \frac{\gamma_1^{(2)} D_2^{(2)} \Delta_2^{(1)}}{2 \pi \Delta_0^{(2)}}, \quad
  q_{21} = -2\beta_2^{(2)} \frac{\gamma_2^{(2)} E_1^{(2)} \Delta_1^{(2)}}{\pi \Delta_0^{(2)}}, \quad
  q_{13} = -2\beta_1^{(2)} \frac{\gamma_1^{(2)} D_3^{(2)} \Delta_3^{(2)}}{\pi \Delta_0^{(2)}}, \\
  q_{31} = -2\beta_3^{(2)} \frac{\gamma_3^{(2)} F_1^{(2)} \Delta_1^{(2)}}{\pi \Delta_0^{(2)}}, \quad
  q_{23} = -2\beta_2^{(2)} \frac{\gamma_2^{(2)} E_3^{(2)} \Delta_3^{(2)}}{\pi \Delta_0^{(2)}}, \quad
  q_{32} = -2\beta_3^{(2)} \frac{\gamma_3^{(2)} F_2^{(2)} \Delta_2^{(2)}}{\pi \Delta_0^{(2)}}.
\end{gather*}

Introducing the notations
\begin{gather*}
  p^-(t) = \begin{cases}
    p(t), & 0 < t < 1\\
    0, & t > 1
  \end{cases}, \qquad
  \psi(t) = f(-t), \qquad
  p_0^-(t) = \begin{cases}
    p_0(t), & 0 < t < 1,\\
    0, & t > 1
  \end{cases},\\
  F^+(t) = \begin{cases}
    0, & 0 < t < 1\\
    \nu_1(t), & t > 1
  \end{cases},\qquad
  D^-(t) = \begin{cases}
    D(t), & 0 < t < 1\\
    0, & t > 1
  \end{cases}\\
  K_1(t,x) = \frac{\lambda_1}{t-x} + \frac{\lambda_2}{t+x} + R_1(t,x), \qquad
  K_2(t,x) = \frac{\lambda_3}{t-x} + \frac{\lambda_4}{t+x} + R_4(t,x)
\end{gather*}
we have the system of integral equations
\begin{gather}
  \frac d{dx} \left(
  \int_0^\infty K_1(t,x) p^-(t) dt
  + \int_0^\infty R_2(-t,x) \psi(t) dt
  \right)
  = \frac 1{D^-(x)} \int_0^x dt\int_0^t [p_0^-(\tau) - p^-(\tau)] d\tau + F^+(x), \qquad x > 0\label{eq:15}\\
  \int_0^\infty K_2(-t,-x) \psi(t) dt + \int_0^\infty R_3(t,-x) p^-(t) dt = q_0(-x), \qquad x > 0\label{eq:16}
\end{gather}
To solve the system (\ref{eq:15})-(\ref{eq:16}), when $D(x) = h_0
x^3$, $x\in(0,1)$, (for example, when $h_1(x)=h_1 x$,
$E_1(x)=E_1=\text{const}$), making the substitution $t=e^\zeta$,
$x=e^\xi$ with notation $\varphi(x) = \int_0^xdt
\int_0^t[p_0^-(\tau)-p^-(\tau)]d\tau$ and using generalized Fourier
transform \cite{23}, we obtain the system
\begin{equation}\label{eq:17}
  \begin{aligned}
    G_1(s) F^-(s) + G_2(s) \Phi(s-2i) &= \Psi^+(s)+P(s)\\
    G_3(s) F^-(s) + G_4(s) \Phi(s-2i) &= Q(s)
  \end{aligned} \qquad -\infty < s < \infty
\end{equation}
where
\begin{gather*}
  F^-(s) = \frac 1{\sqrt{2\pi}} \int_{-\infty}^0 \varphi(e^\zeta) e^{i\zeta s}d\zeta, \qquad
  \Phi(s) = \frac 1{\sqrt{2\pi}} \int_0^\infty \psi(e^\zeta) e^{i\zeta s} d\zeta, \\
  \Psi^+(s) = \frac 1{\sqrt{2\pi}} \int_0^\infty e^{3\xi} F^+(e^\zeta) e^{i\zeta s} d\zeta, \qquad
  P(s) = - \frac{G_1(s) - h_0^{-1}}{s(s-i)} P_1(s-2i), \qquad
  P_1(s) = \frac 1{\sqrt{2\pi}} \int_{-\infty}^0 p_0^-(e^\zeta) e^{i\zeta s} d\zeta, \\
  Q(s) = \frac 1{\sqrt{2\pi}} \int_{-\infty}^0 q_0(-e^\zeta) e^{2\xi} e^{i\zeta s} d\zeta + \frac{i G_3(s)}{s(s-i)} P_1(s-2i), \\
  G_1(s) = -\sqrt{\frac \pi 2} s (s-2i) (s-i) \left[
    \lambda_1 \text{cth}\lambda s + \frac{\lambda_2}{\text{sh} \pi s}
    + \sum_{m\neq n = 1}^3 \frac{\omega_{mn}}{\text{sh}\pi s} \frac{\beta_m^{(1)}}{\beta_n^{(1)2}} \exp\left(
    is \ln\frac{\beta_m^{(1)}}{\beta_n^{(1)}}
    \right)
    \right] - \frac 1{h_0}, \\
  G_2(s) = \sqrt{\frac\pi 2} (s-2i) \sum_{m\neq n = 1}^3 \frac{\alpha_{mn}}{\text{sh}\pi s} \frac{\beta_m^{(1)}}{\beta_n^{(2)2}} \exp\left(
  is \frac{\beta_m^{(1)}}{\beta_n^{(2)}}
  \right), \\ 
  G_3(s) = \sqrt{\frac\pi 2} s(s-i) \sum_{m\neq n = 1}^3 \frac{r_{mn}}{\text{sh}\pi s} \frac{\beta_m^{(2)}}{\beta_n^{(1)2}} \exp\left(
  is \ln\frac{\beta_m^{(2)}}{\beta_n^{(1)}}
  \right), \\
  G_4(s) = i \sqrt{\frac\pi 2} \left[
    \lambda_3 \text{cth} \lambda s
    + \frac{\lambda_4}{\text{sh} \pi s}
    + \sum_{m\neq n = 1}^3 \frac{q_{mn}}{\text{sh} \pi s} \frac{\beta_m^{(2)}}{\beta_n^{(2)2}} \exp\left(
    is \ln\frac{\beta_m^{(2)}}{\beta_n^{(2)}}
    \right)
    \right]
\end{gather*}

Excluding from the system (\ref{eq:17}) function $\Phi(s)$, we obtain
the Riemann problem
\begin{equation}\label{eq:18}
  \begin{gathered}
    \frac{\Psi^+(s)}{\sqrt{s+i}} = \frac{G(s)}{\sqrt{1+s^2}}F^-(s) \sqrt{s-i} + \frac{H(s)}{\sqrt{s+i}}\\
    G(s) = \frac{G_1(s) G_4(s) - G_2(s) G_3(s)}{G_4(s)}, \qquad
    H(s) = \frac{Q(s) G_2(s) - P(s) G_4(s)}{G_4(s)}.
  \end{gathered}
\end{equation}

By virtue of functions $\Psi^+(s)$ and $F^-(s)$ definition, they will
be boundary values of the functions which are holomorphic in the upper
and lower half-planes, respectively.

The problem can be formulated as follows: it is required to obtain the
function $\Psi^+(z)$, holomorphic in the half-plane $\Im z > 0$ and
which vanishes at infinity, and the function $F^-(z)$, holomorphic in
the half-plane $\Im z < 1$ (with the exception of a finite number of
zeros of function $G(z)$) which vanishes at infinity and are
continuous on the real axis by condition (\ref{eq:18}).

Since $\Re G_0(s) > 0$ and $G_0(\infty)=G_0(-\infty)=1$, we have
$\text{Ind} G_0(s) = 0$, $G_0(s)=G(s)/\sqrt{1+s^2}$.

The solution of the problem (\ref{eq:18}) has the form \cite{24}
\begin{equation}\label{eq:19}
  \begin{aligned}
    F^-(z) = \frac{\widetilde X(z)}{\sqrt{z-i}}, \quad
    \Im z \le 0; \qquad
    \Psi^+(z) = \widetilde X(z) \sqrt{z+i}, \quad
    \Im z > 0\\
    F^-(z) = (\Psi^+(z) - H(z)) G^{-1}(z), \qquad 0 < \Im z < 1
  \end{aligned}
\end{equation}
where
\begin{equation*}
  \widetilde X(z) = X(z)
  \left\{
  -\frac 1{2\pi} \int_{-\infty}^\infty \frac{H(t)dt}{X^+(t) \sqrt{t+ i} (t-z)}
  \right\}, \qquad
  X(z) = \exp\left\{\frac 1{2\pi i} \int_{-\infty}^\infty \frac{\ln G_0(t) dt}{t-z}\right\},
\end{equation*}
(here the integral should be understood in the sense of the Cauchy
principal value).

Using the formula $\varphi''(x) = \frac{\varphi_0''(\ln x) -
  \varphi_0'(\ln x)}{x^2}$ and applying the inverse transformation
$\varphi_0'(\ln x) = -\frac i{\sqrt{2\pi}} \int_{-\infty}^\infty s
\Phi^-(s) e^{-is\ln x}ds$, $\varphi_0''(\ln x) = \frac 1{\sqrt{2\pi}}
\int_{-\infty}^\infty s^2 \Phi^-(s) e^{-is\ln x} ds$.
                
We will investigate the behavior of the function
$p_0(x)-p(x)=\varphi''(x)$ in the neighborhood of the points $z=0$ and
$z=1$.

We obtain by an inverse transformation: $p^0(x)-p(x)=O(1)$, $x\to 1-$.

The poles of the function $F^-(z)$ in the domain $D_0=\{z:0<\Im z<1\}$
may be zeros of the function $G(z)$.  It can be shown that the
function $G(z)$ has no zeros in the strip $0 < \Im z < 3/2$.  Then,
applying Cauchy's theorem to the functions $e^{-i\xi z} i z
\Phi^-(z)$, $e^{-i \xi z} z^2 \Phi^-(z)$ we obtain the following
estimate
\begin{equation*}
  p^0(x) - p(x) = O(x^{\delta-2}), \qquad x\to 0+, \quad \delta > \frac 32.
\end{equation*}
Since $\psi(t) = f(-t)$,  crack opening behavior has the form  
\begin{equation*}
  f(x) = O(x^{-1/2+\omega}), \qquad x\to 0-, \quad 0 < \omega < 1/2.
\end{equation*}

\section{Conclusions}

In this paper we consider a piecewise-homogeneous anisotropic plate of
piezoelectric material, weakened by a crack that goes out at the
interface of two materials. The crack propagation is delayed by the
inclusion of an elastic non-homogeneous beam.

The resulting boundary-value contact problem is reduced to a system of
singular integro- differential equations, which is reduced to the
Riemann boundary value problem by the use of integral transformations.

The main result of this paper is that the solution of the problem was
obtained in an explicit form. Also, on the basis of an asymptotic
analysis, it turned out that the normal contact stress along the
contact line of the inclusion with the plate is bounded at one end of
the inclusion. At the other end (cusped and coming out at the
interface of the two materials) of the inclusion the normal contact
stress admits the singularity with order less than 1/2. The order of
the singularity can be also decreased by choosing the geometric and
physical parameters of the problem. At the end of the crack the
singularity of the crack opening function is also decreased under the
action of the inclusion.

The obtained results are significant in the problems of fracture
mechanics and in those of stress concentration. These results can be
successfully applied in geological and geophysical problems,
particularly in the tasks of reinforcement of constructions and rocks
and in delaying of landslide processes.


\providecommand{\bysame}{\leavevmode\hbox to3em{\hrulefill}\thinspace}
\providecommand{\MR}{\relax\ifhmode\unskip\space\fi MR }
\providecommand{\MRhref}[2]{%
  \href{http://www.ams.org/mathscinet-getitem?mr=#1}{#2}
}
\providecommand{\href}[2]{#2}

\end{document}